\documentclass[sigconf]{acmart}

\acmConference[SE2030 '24]{Software Engineering in 2030 Workshop}{July 15--16,
  2024}{Porto Galinhas, BR}
\usepackage{tabularx}
\begin{document}

\title{Addressing Sustainability-IN Software Challenges}


\author{Coral Calero}
\orcid{0000-0003-0728-4176}
\affiliation{%
  \institution{University of Castilla-La Mancha}
  \city{Ciudad Real}
  \country{Spain}
}\email{coral.calero@uclm.es}

\author{Félix O. García}
\orcid{0000-0001-6460-0353}
\affiliation{%
  \institution{University of Castilla-La Mancha}
  \city{Ciudad Real}
  \country{Spain}
} \email{felix.garcia@uclm.es}

\author{Gabriel Alberto García-Mireles}
\orcid{0000-0003-4633-7410}
\affiliation{%
  \institution{University of Sonora}
  \city{Hermosillo}
  \state{Sonora}
  \country{Mexico}
}
\email{gabriel.garcia@unison.mx}

\author{Mª Ángeles Moraga}
\orcid{0000-0001-9165-7144}
\affiliation{%
  \institution{University of Castilla-La Mancha}
  \city{Ciudad Real}
  \country{Spain}
} \email{mariaangeles.moraga@uclm.es}

\author{Aurora Vizcaíno}
\orcid{0000-0002-2072-5581}
\affiliation{%
  \institution{University of Castilla-La Mancha}
  \city{Ciudad Real}
  \country{Spain}
}\email{aurora.vizcaino@uclm.es}

\renewcommand{\shortauthors}{Calero et al.}

\begin{abstract}
In this position paper we address the Software Sustainability from the IN perspective, so that the Software Engineering (SE) community is aware of the need to contribute towards sustainable software companies, which need to adopt a holistic approach to sustainability considering all its dimensions (human, economic and environmental). A series of important challenges to be considered in the coming years are presented, in order that advances in involved SE communities on the subject can be harmonised and used to contribute more effectively to this field of great interest and impact on society. 
\end{abstract}


\begin{CCSXML}
<ccs2012>
   <concept>
       <concept_id>10003456.10003457.10003458.10010921</concept_id>
       <concept_desc>Social and professional topics~Sustainability</concept_desc>
       <concept_significance>500</concept_significance>
       </concept>
   <concept>
       <concept_id>10011007</concept_id>
       <concept_desc>Software and its engineering</concept_desc>
       <concept_significance>500</concept_significance>
       </concept>
 </ccs2012>
\end{CCSXML}

\ccsdesc[500]{Social and professional topics~Sustainability}
\ccsdesc[500]{Software and its engineering}

\keywords{Sustainability IN Software, Challenges}


%\received{--}
%\received[revised]{--}
%\received[accepted]{--}

\maketitle

\section{Introduction}
Software is no longer what it used to be, as it has gone from being a good tool to facilitate the development of some common tasks in the work environment to being a crucial element in our daily lives. Furthermore, the number of software users and their characteristics (differences in age, culture, specialisation, etc.) has also increased and the types of devices on which software runs vary from the most traditional devices (PCs, laptops, mobile phones, etc.) to the most modern ones (household appliances, agriculture machinery, etc.). All these changes have made software a cornerstone of today's society and as software engineers we have to be aware of the enormous impact of all this software and how important it is to work on software sustainability. If software is one of the foundations for the functioning of our society, we have to make sure that it is created in a sustainable way and that it is always done under the guidelines set by the Sustainable Development Goals (SDGs)\cite{SDGs2015}.

The most common definition for sustainability, used in the field of Software Engineering (SE), is related to sustainable development, defined as “development that meets the needs of the present without compromising the ability of future generations to meet their own needs” \cite{Bruntland1987}. Besides, a software is sustainable if it is energy efficient, minimizes the environmental impact of the process it supports, and has a positive impact on social and/or economic sustainability. In this context, software plays a fundamental role, both as part of the problem and as part of the solution, i.e., being the software itself sustainable or achieving sustainability by means of the resulting software within any domain, respectively. These are known as “Sustainability IN Software” and “Sustainability BY Software” \cite{calero2022}. This paper focuses on the IN perspective by identifying a representative set of challenges which can be considered in each one of the affected suistainability dimensions, which are \cite{CaleroMoragaPiattini2021}: 
\begin{itemize}
   \item Economic sustainability: how the software lifecycle processes protect stakeholders’ investments, ensure benefits, reduce risks, and maintain assets.
    \item Human sustainability: how software development and maintenance affect the sociological and psychological aspects of the software development community and its individuals. This encompasses topics such as: labour rights, psychological health, social support, social equity, and liveability.
    \item Environmental sustainability: how software product development, maintenance and use affect energy consumption and the usage of other resources, also usually known as “Green Software”.
\end{itemize}

An alternative approach to characterize sustainability is based on the Karlskrona Manifesto \cite{becker2015} to address the sustainability concerns. The manifesto considers five sustainability dimensions (environmental, social, economic, technical, and individual) \cite{penz2013} and the impact that software has during its development (first order effects), by means its use and application (second order effects) as well as structural changes in society or economy due to software use by long time and by very large number of persons \cite{becker2015}. Its principles address the need to consider systemic effects as well as multiple dimensions, multiple disciplines, and apply long-term thinking. Indeed, Calero et al. \cite{calero2022} noted that the five dimensions of the manifesto are appropriate when software is a component of a system which its purpose is achieving  sustainability goals. Thus, the manifesto content tends towards a sustainability by software perspective and requires expertise in different fields for addressing appropriately sustainability goals. 

From the analysis of 55 systematic literature studies (systematic literature reviews or systematic mapping studies) published between 2010 and 2022, it can be remarked that most of the studies (78{\%}) address the environmental dimension, while economic (7{\%}) and human (11{\%}) dimensions have been little explored and the most common approach is by addressing only one dimension per study (76{\%}). Besides, 69{\%} of studies addressed the sustainability IN software perspective while 15{\%} focused on the sustainability BY software perspective, with the remaining studies considering both perspectives. Furthermore, there is also a general agreement on using the Brundtland definition of sustainable development (15 papers, 27{\%}). Other remarkable point is that dimensions are composed of multiple layers where sustainability  meanings are inconsistent due to the granularity level of the object under study, its context and intervention described in empirical studies\cite{mcguire2023}. Thus, assessment sustainability effects requires to take into account differences between level of granularity (e.g., individual, organisation, industry, municipality, region, country) and consider interactions among several systems \cite{mcguire2023}.

Therefore, facing the challenge of developing sustainable software is a complex endeavor that requires establishing a strategy for understanding better the phenomena to gather new insights, developing methodological support for enhancing the sustainability of software systems and providing appropriate sustainability training. Thus, our position paper centers on the software sustainability from the IN perspective for addressing a set of representative challenges to face in the next years. Section 2 summarises relevant background and previous results in each of the dimensions, to illustrate important challenges to be addressed in the future. Section 3 concludes the paper by identifying some relevant actions of interest that the software engineering community can take into account to address the challenges presented. 

\section{SUSTAINABILITY-IN Dimensions}
\subsection{Economic Dimension}
To the best of our knowledge, the economic sustainability is the least studied dimension from a software perspective. However, it is important to integrate sustainability into the business core and in order to do that, companies should consider sustainability at the level of business governance and strategy. Software companies should ensure that their operations and products are not only economically viable, but also responsible from a human and environmental perspective. The problem is that organisations do not know what actions to take or how to make sustainability part of their business routines and strategies \cite{Baumgartner10}. Moreover, most of the guidelines provide instructions on how to measure and report organisational sustainability performance, but these guidelines lack advice on how to achieve sustainability in the first place \cite{Cleven2012}. 
The application of Enterprise Architecture (EA) fundamentals can be a driver to progress towards achieving a sustainable software enterprise, mainly because sustainability must be incorporated at all organizational levels (strategy and decision-making; business processes; IT portfolio management; information systems design and projects; or technological infrastructure) and EA is useful to manage, in an integrated way, all these different aspects of the organization and its information technology. The initial work relating EA and sustainability is from more than a decade ago but with the focus on sustainable EA by itself \cite{hausman2011sustainable} which has been followed up in works such as  \cite{Perdana20}. This is related with “continuous EA”, an approach that advocates to apply the DevOps continuity principle to EA. A recent paper proposes a Green Enterprise Architecture (GREAN) to integrate Environmental Sustainability into Digital Transformations \cite{Vandevenne23}. Other  related studies mainly address the sustainability and IT by studying sustainability “silos” problem (lack of integrated systems and data) and how EA can help by ensuring alignment between environmental and IT management \cite{Scholtz2014}, or by proposing integrated solutions such as EIRA (European Interoperability Reference Architecture)  which illustrates how the EA artifacts, especially models, can be useful to support the sustainability in all levels of the organizations \cite{Combemale16}. \subsection{Human Dimension}
The report "State of the American workplace" (2023) \cite{Gallup2023} indicates that "only one-third of U.S. employees are engaged in their work and workplace. And only about one in five say their performance is managed in a way that motivates them to do outstanding work". In the case of software engineers, the number of employees experiencing burnout at work is over 80 percent in the United Kingdom. One reason for this could be because software engineers are under great stress since they are confronted with constantly changing user needs and requirements. Indeed, stress and task complexity are important de-motivators of software developers \cite{franca2011}. Besides, a survey of human factors that impact negatively software requirements activities reported that the lack of effective communication or cooperation in the team or with the clients requires more time for conducting the respective tasks \cite{hidellaarachchi2023}. 
Diverse studies indicate that engineers well-being affect their productivity. In particular, the new challenges posed by a global and distributed way of working, including the growing trend towards teleworking where face-to-face communication is reduced, imply the need to analyse more fully and deeply the engineers' stress and motivation factors.
Distributed teams have existed for decades in software development companies. The challenges faced for global software development have been widely analyzed in literature \cite{GRANDE24}. However,  the interest in distributed work increased considerably after the COVID-19 pandemic \cite{SMITE2023JSS,Smite23IEEESW,Ford22} since all the companies were forced to work in a distributed way and new terms appeared such as WFH (Working From Home) when work is conducted from home, WFA (Working From Anywhere) when work is conducted from home or any other alternative location or even Hybrid work when work is interchangeably conducted from the office and remote location \cite{Smite23IEEESW}. 
These new types of work imply new challenges, for instance, in the case of working from home, it increases the feeling of isolation, deteriorates both social ties and team cohesion and decreases interest in collaborative work \cite{Smite23IEEESW}. The factors that affected well-being and productivity during the pandemic were also analyzed by Ralph et al., who conducted a survey that concluded that during the pandemic WFH had a negative effect on productivity and on developers’ wellbeing \cite{Ralph20}. Furthermore, Russo et al. explained how anxiety, distractions or lack of work motivation began to increase during that period \cite{Russo21}. Ozcaya \cite{Ozcaya21} predicted how software engineering would look like after the pandemic and identified the lack of collaboration as a critical challenge as new communication barriers should be analysed. In addition, a hybrid approach involves online communication and coordination, so Ozkaya proposes to reduce dependencies among the tasks carried out by engineers. The lack of informal communication needs also to be addressed in this context as this is an important source for information sharing. Stress, motivation and performance are other fundamental aspects to consider in this context \cite{suarez23}. 
Human sustainability is therefore a key aspect for software organizations, which usually experience employees with low motivation, high rotation rates and, therefore, weak teams. 
Human sustainability encompasses multiple facets such as the training of software professionals, as well as offering them fair and equitable working conditions and opportunities for career advancement with the possibility of access to leadership positions \cite{Elisa24}. In addition to the usual hard and soft skills, it is important to improve the training that companies offer on sustainability issues, and to help from the academic world to train students and companies on the subject \cite{HELDAL24}. 

\subsection{Environmental Dimension}
This dimension, also known as "green", relates to energy consumption and the use of other resources. Studies estimate that the 20{\%} of the world's energy consumption in 2030 will be used by information technologies (IT) \cite{challe6010117}, and an important part of this consumption will be due to software. For as long as computers have existed, software has been evolving, and advances in hardware have also led to more and less profound changes in software evolution. Especially in recent times we are experiencing a major revolution due to three aspects that radically impact the effects that software can have on the environment: the massive use of internet-based software applications, the huge amount of data we generate, and the increasingly presence of artificial intelligence (AI) components in the software being developed.

Software is no longer tied to devices, the evolution of Internet technology has allowed software applications to evolve, giving them the gift of ubiquity, allowing users to access their applications anytime, anywhere and from different devices. According to \cite{digital2024} in its January 2024 report, 66.2{\%} of the population are Internet users with an average usage of 6 hours and 40 minutes per day (users between 16 and 64 years old). Furthermore, 61.8{\%} of these users access the internet from a laptop or desktop, while 94.6{\%} access it from a smartphone. The ubiquity of software made it possible, for example during the COVID-19 pandemic, for much of the business activity to be moved to employees' homes and to remain active, as also stated in the previous subsection. But this also affected to the personal leisure activities and the use of social networks is a good example of Internet-based software applications which can have high impact on this dimension. In \cite{digital2024} it is estimated that there are 5.04 billion social media users.

In terms of data, the evolution of data centres has brought many benefits, such as freeing users from the need to store and manage their data, but it has also made them less aware of the amount of data they generate, so data centres have become behemoths that need to be able to store more and more data. According to \cite{racounter2019}, accumulated digital data grew from 4.4 zetabytes in 2019 to 44 zetabytes in 2020. And this trend seems set to continue or even worsen over time, with 463 exabytes of data expected to be created every day by 2025 \cite{racounter2019}. All this data centers load has led to estimates that (in the expected scenario) electricity consumption will triple in 7 years, from 1,000 TWh in 2023 to almost 3,000 TWh in 2030 (four times more in the worst case scenario) \cite{challe6010117}. Data is not only generated at a business level but also at a personal level, today the cloud is used as the main storage medium, which is provided free of charge (at least up to a certain amount) to all users.

Finally, AI has been identified globally as one of the technologies with the greatest projection and impact in all areas of activity. AI is having a strong transformative impact on multiple sectors of activity. For example, it is estimated that the application of data and artificial intelligence in Spanish industry will have an estimated impact on GDP of €16.5 billion by 2025. Tech giants like Google, Apple, Microsoft and Amazon spend billions to create those products and service and companies are spending nearly 20 billion dollars on AI products and services. According to a 2023 IBM survey, 42\% of enterprise-scale businesses integrated AI into their operations, and 40\% are considering AI for their organizations. In addition, 38\% of organizations have implemented generative AI into their workflows while 42\% are considering doing so \cite{thomas2024}. And one of the issues which emerged with the advent of AI is how much energy it consumes. \cite{strubell2019energy} estimates that training computers to learn human language produces five times the CO2 emissions of a car over its lifetime. Or in \cite{Gutirrez2023GreenINML}, where the authors illustrate the need to achieve a good trade-off between accuracy and energy efficiency of models, since an improvement of only 0.02{\%} in accuracy implies a doubling of energy consumption, while the model with the best accuracy consumes 30 times more energy than the most energy-efficient model, with a difference of 9{\%} in accuracy. Moreover, it is remarkable that nowadays AI tools are widely and easily accessible by users to solve problems in their work and personal lives, as is the case of well-known examples such as Copilot, Chat-GPT and similar ones. This, on the other hand, can have a great impact on this dimension if this technology is not developed and used responsibly.

We can therefore metaphorically refer to a  three-headed hydra of software  consumption that we must at least be able to control, as it is expected to be very difficult to overcome. 
\section{Fostering Sustainability-IN Software: Future Directions}
Based on the previous findings, we identify several research lines which require further action to foster software Sustainability-IN perspective.
\newline
\noindent \textbf{General}
\begin{itemize}
\item \textit{Harmonizing sustainability vocabulary in SE.} Until now, there is a lack of agreement as regards what is the definition of sustainable software and how it can be measured \cite{venters2023}. About sustainability dimensions, research have been focused on the environmental concerns, particularly studying how to reduce energy consumption of software \cite{calero2017}. Thus, developing an agreed ontology can help to organize the sustainability terms and provide precise definition of concepts.
\item \textit{Holistic approach for Sustainability}. Sustainability is a key challenge for software organizations, which requires developing new models, frameworks, practices, and technologies in different areas. This effort to address the sustainability of software organizations in such a comprehensive way requires the combination of knowledge and experience in many research areas of software engineering and an holistic view to properly integrate the contributions in the different involved dimensions is required and manage the interactions between them. An holistic approach also involves  collaborating of Software Engineering field with other disciplines, such as economy, psychology, among others.
\item \textit{Developing sustainability assessment frameworks for addressing sustainability dimensions and impacts.} Organisations has problems for defining relevant sustainability measures \cite{heldal2024}. We can rely on the quality in use \cite{iso25010}, which characterizes the effect that software has on stakeholders, considering the characteristics of the users, social environment and tasks. Software quality models could support the development of domain ontologies for assessing software impacts considering relevant indicators and factors for the sustainability dimension on study. For the environmental  dimension, speficic ontologies for measuring the energy impact of software can serve as starting point \cite{Mancebo2021}.
\end{itemize}
\noindent \textbf{Economic}
\begin{itemize}
    \item \textit{Application of Enterprise Architecture (EA) fundamentals to achieve a sustainable enterprise.} It is necessary to develop an enterprise architecture framework, including techniques, viewpoints and types of models, to support the implementation of sustainability improvement within the organisation. 
    \item \textit{Definition of sustainable business and maturity models.} Software development companies should innovate their business models to become more sustainable. Moreover a specific Sustainability-IN reference model which include the required actions per each dimension and organised by maturity and capability levels can better guide software organization to organize its improvement efforts in this complex field. As an input for this reference model from economic dimension view, software companies should include specific actions to not only achieve an economic benefit but also consider  sustainability aspects, as for instance: 
    \begin{itemize} 
    \item Work with suppliers who have sustainable practices without considering the cost. 
    \item Software solutions must ensure compliance with country-specific legislation (such as GDPR) to avoid the costs of non-compliance. 
    \item When possible foster: repairing instead of replacing hardware; open-source code instead of licensed software
    \item Software operations should be done according to the balance between costs and registered sustainability assessment criteria. 
    \item The security and privacy of business and customer data must be ensured to avoid excessive costs due to threats to data. 
    \item Customers must be provided with IT solutions that optimise resources, minimising unnecessary expenses. 
    \end{itemize} 
\end{itemize} 
\noindent \textbf{Human}
\begin{itemize}
 \item \textit{Considering the expertise of specialists on human and social behavior}. Topics such as motivation of software developers have been addressed in the SE field as well as soft-skills. However, the human dimension describes the need to consider sociological and psychological aspects which address topics such as labour rights, psychological health, or social equity, among others. Thus, a multidisciplinary approach can provide new insights for addressing the challenges that software developers faces on new work modalities.
 \item \textit{Define software sustainability training programmes for employees.} Company employees should have available specific training courses on sustainability as well as software sustainability reference materials containing good practices or recommendations.
 \item \textit{Providing career development opportunities for employees.} The company should facilitate the career development of its employees, ensuring internal promotion and encouraging access to leadership positions within the company. 
\item \textit{Human capital management in software companies must promote open working environments}. These must be characterized by trust, mutual respect and empathy, fostering formal and informal knowledge sharing. Different forms of work, whether home-based, remote or hybrid, should be taken into account in order to optimise their  positive impact on the human dimension.   
\end{itemize}
\noindent \textbf{Environmental}
\begin{itemize}
\item \textit{The use of resources must be a primary consideration in software development}. If at the beginning it was the \textbf{what} that mattered (the advantage was in the automation of functionalities), then the \textbf{how} (quality became the differentiating element), we are now in the era of the \textbf{with what} (the resources used during development and use must be considered as elements of the first importance). 
\item \textit{A robust body of empirical knowledge must be built up from good practices identified along the software life cycle.}
\item \textit{It is essential to provide software professionals with the required support.} Based on the inputs from the former point, tools, guidelines, recommendations and documentation are needed to help and facilitate the integration of environmental aspects in the development of software.  
\item \textit{As software is constantly evolving, everything related to its resource consumption must evolve with it.} Resource usage must therefore be kept under constant review and improvement.
\item \textit{End-users have the right to know the environmental impact of software solutions available on the market.} This will enable them to make informed choices.
\item \textit{Governments need to move in the same direction.} Legislation to support the development of green software is needed.  
\item \textit{Proactively monitor the contributions that disruptive technologies can make for obtaining greener solutions.} A good example of this nowadays is the possibilities that quantum computing offers for optimising certain types of problems that require a high computational load on classical computers. 
\end{itemize}
These and other actions should be part of the future of sustainable software development.

\section*{Acknowledgments}
The work has received support from the following projects: OASSIS (PID2021-122554OB-C31/ AEI/10.13039/501100011033/FEDER, UE); EMMA (Project SBPLY/21/180501/000115, funded by CECD (JCCM) and FEDER funds)

\bibliographystyle{ACM-Reference-Format}
\bibliography{paperSE2030}

\end{document}